\documentclass[paper,aps,12pt,floats,tightenlines]{revtex4}
\usepackage{amssymb}
\newcommand{\bld}[1]{\mbox{\boldmath{$#1$}}}

\newcommand{\beg}{\begin{equation}\label}
\newcommand{\en}{\end{equation}}

\begin{document}
\title{\vskip-3cm{\baselineskip14pt \begin{flushright}
\normalsize TTP/05-22\\ \normalsize Nov. 2005\\
\end{flushright}} \vskip0.5cm

On the spectra of atoms and hadrons}
\author{Hartmut Pilkuhn}
\affiliation{Institut f\"ur Theoretische Teilchenphysik, Universit\"at,
D-76128 Karlsruhe, Germany}

\begin{abstract}{For relativistic closed systems, an operator is
explained which has as stationary eigenvalues the squares of the total
cms energies, while the wave function has only half as many components
as the corresponding Dirac wave function. The operator's time
dependence is generalized to a Klein-Gordon equation. It ensures
relativistic kinematics in radiative decays. The new operator is not
hermitian.} \end{abstract}
\maketitle
\vspace{0.8cm}
Energy levels of bound states are calculated by a variety of methods,
which include relativity at least approximately. In atomic theory, the equation
$H\psi_D=E\psi_D$ is used, where $H$ is the $n$-body Dirac-Breit
Hamiltonian, and $\psi_D$ is an $n$-electron Dirac spinor with $4^n$
components. For half a century, 
great hopes were attached to the Bethe-Salpeter equation, for example
in the calculation of positronium spectra \cite{BS}. Most of these
methods have been adapted to hadrons, namely to mesons as quark-antiquark
bound states and baryons as three-quark bound states. New methods such
as nonrelativistic quantum electro-dynamics (NRQED) and numerical
calculations on a space-time lattice have been added. In this note,
the recent extension of another new method to radiative decays is presented.
Its time-independent form is similar to $H\psi_D=E\psi_D$, namely
$M^2\psi=E^2\psi$, but $\psi$ has only half as many components as $\psi_D$.
It applies only to closed systems, where $E$ denotes the total cms energy.
For atoms, this implies a relativistic inclusion of the nucleus, which
is of little practical importance. It is relativistically exact, which
does not mean that it is exact. Rather, relativity ensures
a symmetry which is lost if only one of the constituents is treated
nonrelativistically. It is most important for the lighter
mesons and baryons which have so far resisted any quantitative treatment.
If one will ever find a reliable $M^2$ for mesons and baryons, it will
automatically contain the relativistic kinematics of radiative decays.
This kinemetics is very simple: In the emission of a photon of energy
$\hbar\omega$, momentum conservation endows the system with a recoil
momentum $\bld P=-\hbar\bld k$ opposite to the photon momentum, with
$k=\omega/c$. The resulting photon energy is
\beg{1}\hbar\omega=(E^2-E'{}^2)/2E,\en 
where $E'$ is the total energy of the final state in its own cms.
The photon spectrum calculated from Dirac's time-dependent
perturbation theory, $i\hbar\partial\psi_D/\partial t=H\psi_D$,
disagrees with (\ref{1}), with the single exception of the radiation
from Landau levels in a constant magnetic field. The amount of
disagreement depends on the number of relativistic recoil corrections
in $H$. Via dispersion relations, the photon spectrum also affects the
Lamb shift of energy levels.

For two-body atoms such as muonium (e$^-\mu^+$) or positronium (e$^-$e$^+$),
$M^2$ was first derived approximately from the Dirac-Breit equation \cite{MP},
later directly from the two-body S-matrix of perturbative QED \cite{HHP}.
It is discussed extensively in \cite{P}, with the notation
$M^2=2h$. For the eigenvalues of $M^2$,
particle theorists sometimes use the symbol of one of the three
Mandelstam variables, $s=E^2$. The $E^2$ -form has recently been
derived from the Dirac-Breit equation for arbitrary $n$ \cite{P05}.

In the following, units $\hbar=c=1$ will be used, and
$i\hbar\partial/\partial t$ will be abbreviated as $\partial_t$.
The time-dependent version of $M^2$ was first found empirically by
noting that 
\beg{2}i\partial_\tau\psi(\tau)=M^2(\tau)\psi(\tau),\quad\tau=t/E\en
does reproduce (\ref{1}) \cite{P04}.
Trusting Lorentz invariance, this equation is extended in a
forth-coming review \cite{CP} to a Klein-Gordon equation,
\beg{3}\square\psi=M^2\psi,\quad\square=\nabla^2-\partial_{t,lab}^2,\en
where $t_{lab}$ is the lab time of the moving atom or hadron. The
eigenvalue of $\nabla^2$ is $-\bld P^2$, which by definition vanishes
in the cms (the atomic rest system). There, one has $t_{lab}=t,\;
-\partial_t^2\psi=M^2\psi$. In first-order perturbation theory,
(\ref{2}) follows from (\ref{3}) by the ansatz $\psi=exp\{-iEt\}\psi(\tau)$.
For a free stable atom or hadron of energy $E_{lab}$ and momentum
$\bld P$, (\ref{3}) is the quantum version of the Einstein relation,
\beg{4}E_{lab}^2-c^2\bld P^2=E^2\equiv M^2c^4.\en
Here $c$ is re-inserted in order to display the form $E=Mc^2$, even
though the relevant operator is $M^2$. It is the mass-square
operator. In relativity, the mass itself appears only in
nonrelativistic expansions. 

The essential point is of course to construct $M^2$. However, before
going into details, it may be adequate to mention a property which is
unusual and even embarrassing, at least for physicists: $M^2$ is not hermitian.
There are in fact two equivalent stationary versions,
\beg{5}M^2\psi=E^2\psi,\quad (M^2)^\dagger\chi=E^2\chi.\en
It is of course known that also non-hermitian operators may have real
eigenvalues, but in the past this has been regarded as an avoidable
complication. Decomposing $M^2$ into its hermitian and antihermitian
components, 
\beg{6}M^2=M^2_h+M^2_a,\quad (M^2)^\dagger=M^2_h-M^2_a,\en
it is clear that the real eigenvalue $E^2$ cannot depend on the sign
of $M^2_a$. If one of the two equations applies, the other applies as
well.

The construction of $M^2$ is best explained by using Dirac spinors,
with a separate set of Dirac gamma-matrices for  each particle (the
nucleus is also treated like a Dirac particle). From the individual chiral
matrices $\gamma^5_i$ (with eigenvalues $\pm1$), one constructs a total
chirality matrix, $\gamma^5_{tot}=\gamma^5_1\dots\gamma^5_n$, which
also has eigenvalues $\pm1$. The corresponding components of $\psi_D$
are called $\psi$ and $\chi$, respectively:
\beg{7}\gamma^5_{tot}\psi=\psi,\quad\gamma^5_{tot}\chi=-\chi.\en
As $\gamma^5_{tot}$ is not conserved, the Dirac-Breit equation couples
components of opposite total chiralities, but one component is easily
eliminated in terms of the other, resulting in (\ref{5}). Obviously,
the elimination of components turns the original hermitian operator
into a nonhermitian one in the subspace.

The appearance of $E^2$ follows from a coordinate transformation and
the subsequent elimination of components. The substitutions $\bld
r_i=E\bld r_{i,E}$ and the definition $r_{ij} =|\bld r_i-\bld r_j|$
allow one to rewrite $V_{ij}=q_iq_j/r_{ij}=EV_{ij,E}$, and
correspondingly for the momentum operators. A factor $E$
may be extracted from the whole Dirac-Breit operator (provided one
refrains from ``positive-energy projectors''). Elimination of
half of the components transforms the factor $E$ into $E^2$.

For two fermions, the approximate construction of $M^2$ from the
Dirac-Breit operator leads to amazing cancellations. The Dirac-Breit
operator is linear in the particle momenta $\bld p_1$ and $\bld
p_2$. With $\bld p_1=-\bld p_2\equiv \bld p$ in the cms, the expected
bilinear terms cancel out. A somewhat tricky transformation $\psi=C_1\psi_1$
eliminates the kinetic energy of particle 2 altogether. The final
result is an effective Dirac equation similar to that of hydrogen,
with four components for the effective electron, and another factor of
two for the hyperfine interaction from the spin of the effective
proton. It remains valid for positronium, apart from the virtual
annihilation interaction.

In general, eigenstates of chirality are not parity eigenstates. The
parity transformation contains the product of the Dirac matrices
$\gamma^0_i=\beta_i$, $\beta_{tot}=\beta_1\dots\beta_n$. For even $n$,
however, $\beta_{tot}$ commutes with $\gamma^5_{tot}$, and common
eigenstates do exist. In this case, each equation (\ref{5}) is
separately parity invariant. For $n=2$, the nonhermiticity of $M^2$ is
restricted to the hyperfine operator, which also contains a factor $E^{-2}$.

Other operators expected from the Dirac-Breit equation for $n=2$
are $V^2$ and
retardation operators, which are part of the Breit operators. They all
disappear from $M^2$. A more precise derivation of $M^2$ uses the S-matrix
of QED, which must be reduced from its standard $16\times16$-form to $8\times8$
before the interaction is abstracted as the Fourier transform of the
$T$-matrix. If one restricts oneself to the one-photon exchange, the
absence of $V^2$ is trivial. If on the other hand one works with the
traditional 16-component formalism, one also obtains an operator $V^2$
at some stage of the calculation. This operator is cancelled by a
two-photon exchange contribution at a later stage.

When the S-matrix for three fermions is constructed from two-fermion
submatrices, one finds that also the case $n=3$ has some
cancellations. Otherwise, $n=3$ is infinitely more complicated
than $n=2$. The cms condition $\bld p_1+\bld p_2+\bld p_3=0$ is not
nearly as helpful as $\bld p_1+\bld p_2=0$.
$\beta_{tot}$ and $\gamma^5_{tot}$ anticommute for $n=3$, and
$\beta_{tot}$ exchanges $\psi$ with $\chi$. This creates a rather
strange situation: The two separate equations (\ref{5}) are coupled by the
parity transformation and by nothing else.

The absence of odd powers of $E$ can be traced back to the CPT theorem
of quantum field theory. According to Feynman, a plane wave
$exp\{-ik_0t+ i\bld{kr}\}$ is associated not only with a particle of
energy $k_0$ moving forward in time, but also with an antiparticle of
energy $-k_0$ moving backwards. A formalism which reproduces the
S-matrix of QED and QCD includes automatically antiatoms and
antihadrons as solutions of energies $-E$ of closed systems.

\end{document}